\documentclass[nofootinbib,reprint,superscriptaddress,preprintnumbers]{revtex4-2}

\makeatletter 
    
\renewcommand\onecolumngrid{
\do@columngrid{one}{\@ne}%
\def\set@footnotewidth{\onecolumngrid}
\def\footnoterule{\kern-6pt\hrule width 1.5in\kern6pt}%
}

\renewcommand\twocolumngrid{
        \def\footnoterule{
        \dimen@\skip\footins\divide\dimen@\thr@@
        \kern-\dimen@\hrule width.5in\kern\dimen@}
        \do@columngrid{mlt}{\tw@}
}%

\makeatother    

\usepackage[bottom]{footmisc}
\usepackage{amsmath,amssymb,marginnote}
\usepackage{multirow} 
\usepackage{booktabs} 
\usepackage{graphicx,subfigure}
\usepackage{dcolumn}
\usepackage{bm}
\usepackage[mathlines]{lineno}
\usepackage[colorlinks=true, allcolors=blue]{hyperref}
\usepackage[usenames]{color}
\usepackage{xcolor}

\begin{document}

\preprint{LA-UR-25-22643}

\title{Asymptotic states of fast neutrino-flavor conversions in the three-flavor framework}

\author{Jiabao Liu}
\affiliation{Department of Physics and Applied Physics, School of Advanced Science \& Engineering, Waseda University, Tokyo 169-8555, Japan}
\author{Hiroki Nagakura}
\affiliation{Division of Science, National Astronomical Observatory of Japan, 2-21-1 Osawa, Mitaka, Tokyo 181-8588, Japan}
\author{Masamichi Zaizen}
\affiliation{Department of Earth Science and Astronomy, The University of Tokyo, Tokyo 153-8902, Japan}
\author{Lucas Johns}
\affiliation{Theoretical Division, Los Alamos National Laboratory, Los Alamos, NM 87545, USA}
\author{Shoichi Yamada}
\affiliation{Department of Physics, School of Advanced Science \& Engineering, Waseda University, Tokyo 169-8555, Japan}
\affiliation{Research Institute for Science and Engineering, Waseda University, Tokyo 169-8555, Japan}

\begin{abstract}
There has been growing evidence that mu- and tau neutrinos are noticeably different due to the emergence of muons in core-collapse supernovae (CCSNe) and binary neutron star mergers (BNSMs). Recent theoretical studies also suggest that all flavors of neutrinos and antineutrinos inevitably experience some flavor mixing instabilities including fast neutrino flavor conversions (FFC), which corresponds to one of the collective neutrino oscillations powered by neutrino self-interactions. This represents a need for quantum kinetic treatment in the numerical modeling of neutrino dynamics, which is, however, a formidable computational challenge. In this paper, we present an approximate method to predict asymptotic states of FFC without solving a quantum kinetic equation under a three-flavor framework, in which mu- and tau neutrino distributions are not necessarily identical to each other. The approximate method is developed based on a Bhatnagar–Gross–Krook (BGK) relaxation time prescription, capable of capturing essential features of mixing competitions among three different flavor-coherent states. Our proposed scheme is computationally inexpensive and easy to implement in any classical neutrino transport scheme.
\end{abstract}
\maketitle

\section{Introduction}
Understanding inner dynamics of core-collapse supernova (CCSN) and binary neutron star merger (BNSM) demands sophisticated numerical simulations due to the complex interplay of micro- and macroscale physical processes (see recent reviews, e.g., \cite{janka2025longtermmultidimensionalmodelscorecollapse,YAMADA2024pjab.100.015,FISCHER2024104107,burrows2021core,mezzacappa2020physical}). The vast disparity in temporal and spatial scales among different physical processes renders direct first-principles simulations computationally prohibitive. This means that a coarse-grained approach is indispensable for numerical modeling, while the challenge lies in developing approximate methods that ensure physical fidelity.

At the moment, this issue is particularly represented by neutrino quantum kinetics (see, e.g., \cite{RevModPhys.96.025004, Richers2020, universe8020094, annurev:/content/journals/10.1146/annurev-nucl-102920-050505,lucas2025} for recent reviews). In CCSN and BNSM environments, neutrino flavor conversions driven by collective interactions of neutrinos themselves have the potential to affect the neutrino radiation field radically, thereby being a game-changing piece of physics to substantially alter nucleosynthesis, shock dynamics, and remnant evolution from those proposed in the literature. Quantifying the effects of neutrino flavor conversions in CCSN and BNSM theories requires quantum kinetic treatment of neutrino dynamics. However, implementing quantum kinetic neutrino transport in CCSN and BNSM simulations presents a significant challenge. The primary one is a scale separation. Resolving the microscopic quantum kinetics of neutrinos alongside macroscopic hydrodynamics is intractable (see, e.g., \cite{PhysRevLett.129.261101,PhysRevD.107.083016,PhysRevD.111.043028} ). Because of the technical challenge, existing strategies to incorporate flavor conversions in CCSN \cite{PhysRevD.107.103034,PhysRevLett.131.061401,10.1093/ptep/ptac118,wang2025effectfastflavorinstabilitycorecollapse} and BNSM \cite{PhysRevLett.126.251101,PhysRevD.106.103003,PhysRevD.105.083024} simulations, by contrast, often depend on phenomenological mixing schemes. While pragmatic, these approaches lack a rigorous connection to quantum kinetic theory, potentially leading to unphysical outcomes even qualitatively. Addressing the issue requires the development of a subgrid model firmly grounded in neutrino quantum kinetics.

Thus far, much effort has been dedicated to developing accurate subgrid models of neutrino flavor conversions. One of the most important tasks is to determine asymptotic states of flavor instabilities. Flavor conversions associated with these instabilities occur in short spatiotemporal scales compared to other physical ones in CCSN and BNSM dynamics, indicating that the neutrino radiation field would achieve a local quasi-steady state. This suggests that the effects of flavor conversions can be included if we can predict such a local quasi-steady (or asymptotic) state. Recently, there have been several attempts to determine them by various ways for fast neutrino-flavor conversion (FFC) in \cite{PhysRevD.106.103039,PhysRevD.107.103022,PhysRevD.107.123021,PhysRevD.108.063003,PhysRevD.109.043024,PhysRevD.110.123018,PhysRevD.110.103019}, and collisional flavor instabilities \cite{zaizen2025spectraldiversitycollisionalneutrinoflavor}. In our previous paper \cite{PhysRevD.109.083013}, we demonstrated that these approximate schemes for determining asymptotic states of flavor conversions can be implemented in classical neutrino transport simulation based on a Bhatnagar–Gross–Krook (BGK) relaxation time prescription (see also \cite{PhysRev.94.511}). This represents a great potential for the effects of flavor conversions associated with collective neutrino oscillations to be incorporated in any classical neutrino transport methods, which paves the way for more accurate CCSN and BNSM simulations with flavor conversions.

However, there is an important limitation in these methods, which can be seen in our previous study \cite{PhysRevD.107.103022}. We provided an analytic prescription to determine asymptotic states of FFC in both two- and three-flavor systems. In the three flavor case, however, $\mu$- and $\tau$ neutrinos (hereafter they are denoted as $\nu_{\mu}$ and $\nu_{\tau}$, respectively) are not distinguished but rather treated collectively. Although it has been believed that this is a reasonable approximation for CCSN and BNSM neutrino radiation field, some recent studies have shown that noticeable differences may appear among $\nu_{\mu}$, $\nu_{\tau}$, and their antipartners, due to muon creations around hot neutron stars \cite{PhysRevLett.119.242702,PhysRevD.102.123001,gieg2024rolemuonsbinaryneutron,refId0,pajkos2024influencemuonspionstrapped,ng2024accuratemuonicinteractionsneutronstar}. Also, the vacuum term with matter-suppressed mixing angles may differentiate $\nu_{\mu}$ and $\nu_{\tau}$ at the initial perturbed state due to the mass-squared difference \cite{shalgar2025neutrinoquantumkineticsflavors}. These suggest that the analytic scheme proposed in \cite{PhysRevD.107.103022} and some relevant methods in \cite{PhysRevD.108.063003,PhysRevD.110.123018} need to be improved. Addressing this issue is, however, not an easy task, that motivates the present work.

This paper is structured as follows. In Sec.~\ref{sec:nontrivial}, we first describe difficulties in predicting asymptotic states of FFC in a three-flavor framework, which underlines the importance of the present work. We then propose a BGK prescription to address the issue in Sec.~\ref{sec:BGKmethod}, and its capability is tested by comparing it to quantum kinetic neutrino transport simulations in Sec.~\ref{sec:validation}. After we provide discussions of how the proposed method is implemented in classical neutrino transport as a subgrid model of FFC in Sec.~\ref{sec:impsubgrid}, we summarize our conclusions in Sec.~\ref{sec:conclusions}. Throughout this paper, we use natural units ($c=\hbar=1$) and the (-+++) metric signature.

\section{Complex interplay among multiple coherent states}\label{sec:nontrivial}

\begin{figure}
    \centering
    \includegraphics[width=1.0\linewidth]{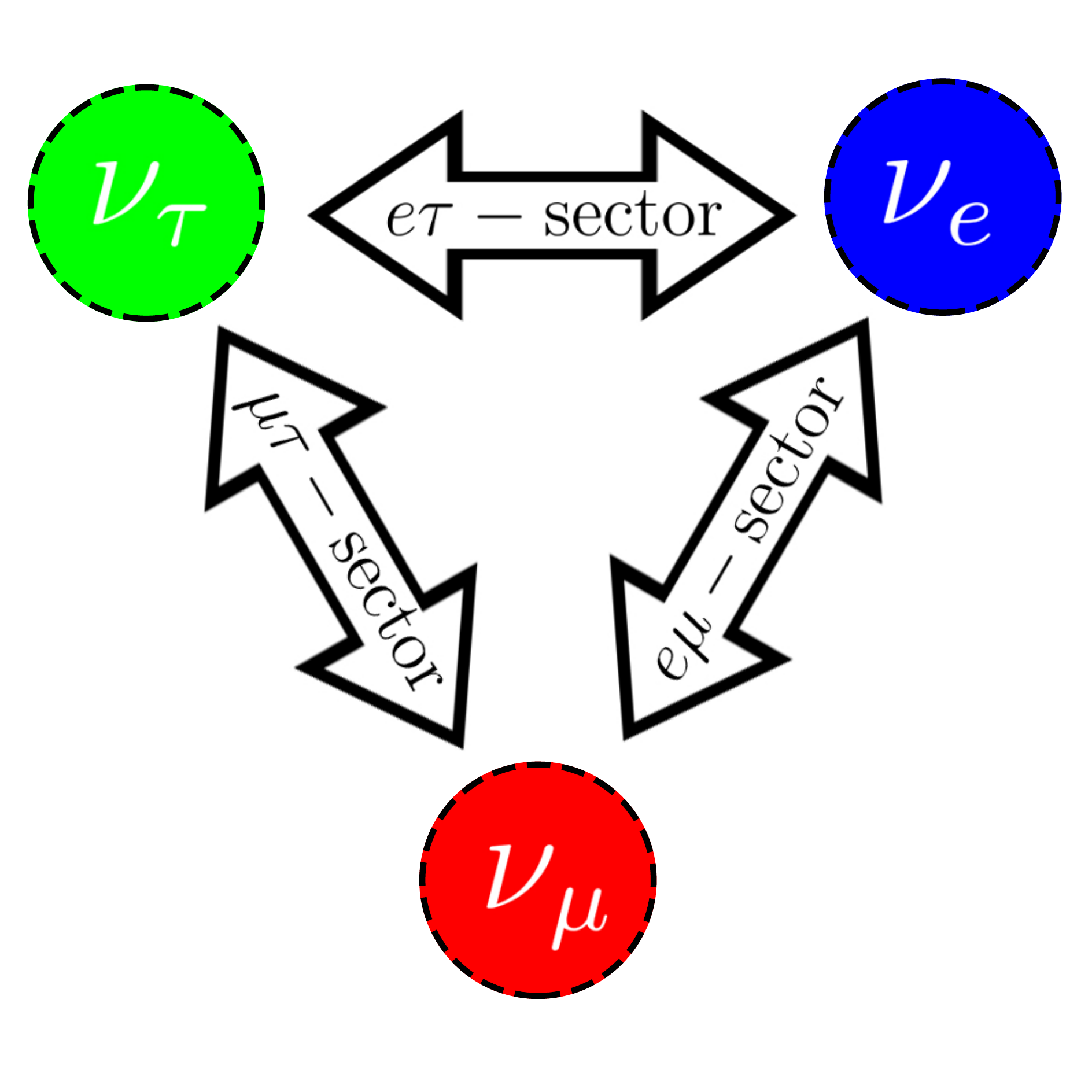}
    \caption{Schematic picture for flavor coherency in three flavor systems. Filled circles represent neutrino flavor eigenstates, where colors distinguish their flavors. The arrows connecting two flavor states represent a flavor coherent state. See text for more details.
    }
    \label{fig:scheme}
\end{figure}

\begin{figure}
    \centering
    \includegraphics[width=1.0\linewidth]{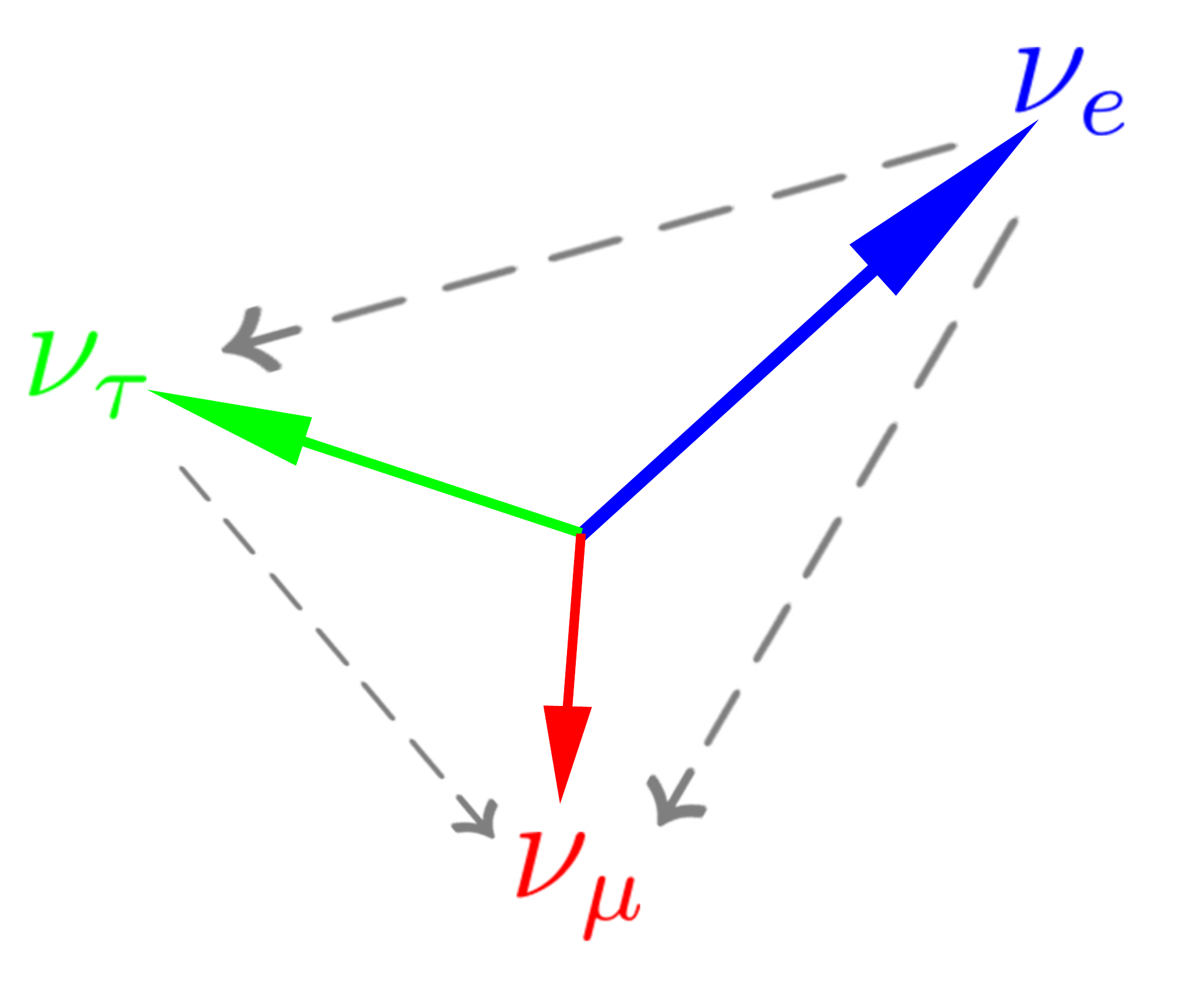}
    \caption{
    An example of the competing effect in three-flavor neutrino flavor conversions. The length of an arrow with a solid line represents occupation numbers of each flavor of neutrinos at given a neutrino angle and energy, where colors distinguish flavors (which is the same as Fig.~\ref{fig:scheme}). In this example, the occupation number of each flavor is assumed to be $\nu_e > \nu_{\tau} > \nu_{\mu}$ and we also assume that flavor conversions make the system flavor equipartition (although it is not always true for FFC; see \cite{PhysRevD.109.083031}). Dashed arrows portray a direction of neutrino flavor conversion. A competing effect can be seen between two flavor coherency associated with flavor conversions in $\nu_{\tau}$. FFCs in $\nu_{e}-\nu_{\tau}$ sector increase $\nu_{\tau}$, but those in $\nu_{\mu}-\nu_{\tau}$ make $\nu_{\tau}$ reduce. See text for more details.}
    
    \label{fig:3fpic}
\end{figure}

In the three-flavor system, there are three independent coherent states ($\nu_e-\nu_{\mu}$, $\nu_e-\nu_{\tau}$, and $\nu_{\mu}-\nu_{\tau}$, see Fig.~\ref{fig:scheme}). For cases where flavor coherency is weak, the associated flavor instabilities occur independently in each coherent sector. This indicates that determining the dynamics of flavor conversion is essentially the same as that in the two flavor cases, which is simple and straightforward. However, after the system enters into a nonlinear phase where the flavor coherency grows large enough to affect the neutrino background, three different coherent states influence each other. The nonlinear saturation of flavor conversions would be regulated by a balance between different coherent states.

In Fig.~\ref{fig:3fpic}, we provide an example where the time evolution of flavor conversions in three flavor systems becomes complicated. In this example, we consider the case that the distribution function of neutrinos at a given neutrino angle and energy has a relation of $\nu_{e} > \nu_{\tau} > \nu_{\mu}$, and we assume that flavor conversions make the system flavor equipartition. For $\nu_{e}$ and $\nu_{\mu}$, we can easily predict their time evolutions; the former and latter should decrease and increase with time, respectively. However, the time evolution of $\nu_{\tau}$ is determined by the balance of the strength of flavor conversion between two flavor coherency, in which $\nu_{\tau}$ gains neutrinos from $\nu_e$ but it converts $\nu_{\mu}$, exhibiting a competing effect. As we shall show below, the strength of flavor conversion is, in general, characterized by multiple factors, which need to be taken into account appropriately.

Here, let us provide another example to illustrate the difficulty in predicting asymptotic states of flavor conversions. As an initial condition, we set neutrino distributions of three flavors so that $\nu_e-\nu_{\tau}$ and $\nu_{\mu}-\nu_{\tau}$ sectors are stable but $\nu_e-\nu_{\mu}$ is unstable. In the early phase, the $\nu_e-\nu_{\mu}$ coherency grows exponentially, and then the flavor conversion affects both $\nu_e$ and $\nu_{\mu}$ distributions once the instability reaches the nonlinear regime. Since $\nu_e$ and $\nu_{\mu}$ distributions are changed, other sectors ($\nu_e-\nu_{\tau}$ and $\nu_{\mu}-\nu_{\tau}$) could be unstable in flavor instability. This triggers a complex interplay of flavor conversions among different coherent sectors. This example exhibits that we cannot determine which coherent sector becomes unstable during the time evolution of the system solely based on the linear stability analysis of initial conditions. The competing effects among different neutrino flavor coherency are also time-dependent, which needs to be modeled one way or another. This argument illustrates that predicting asymptotic states of FFC in three-flavor systems requires a substantial change of approximate scheme from those used in the literature. To this end, we make the best use of BGK prescription, whose detailed formulation is presented in the following section.

\section{BGK method to predict asymptotic states of FFC} \label{sec:BGKmethod}

\subsection{Basic idea} \label{subsec:basicidea}
We start with a brief review of our BGK subgrid model presented in \cite{PhysRevD.109.083013}, which provides a fundamental concept to predict asymptotic states of flavor conversions in the three flavor system. In this subgrid approach, we approximate the dynamics of flavor conversion as a relaxation process. More specifically, the following quantum kinetic equation of neutrinos,
\begin{align}
    \left(\partial_t+\boldsymbol{v}\cdot\nabla\right) f = -i [H,f],
    \label{eq:QKE_flat}
\end{align}
is reformulated as
\begin{align}
    \left(\partial_t+\boldsymbol{v}\cdot\nabla\right) f = - \frac{1}{\tau} ( f - f^a ),
    \label{eq:BGK_flat}
\end{align}
where $f$ denotes the density matrix of neutrinos, and $t$ and $\boldsymbol{v}$ represent the time and three-dimensional spatial unit vector specifying the neutrino flight direction. In the right-hand side of Eq.~\ref{eq:BGK_flat}, $\tau$ and $f^a$ denote a relaxation time and asymptotic state of $f$ due to flavor conversions, respectively. One of the noticeable properties in the BGK formalism is that the strength of flavor conversion can be specified by the relaxation time, which is a useful property to handle competing effects of flavor conversions when there are more than one flavor sectors (see below).

For FFC dynamics in two flavors (or $\nu_{\mu}=\nu_{\tau}$, i.e, only one flavor sector exists) systems with periodic boundary conditions in space, we can roughly estimate $\tau$ by an empirical relation of growth rate of FFIs \cite{Nagakura_2019,PhysRevResearch.2.012046}, while the asymptotic state ($f^a$) can be approximated by using an analytic scheme of \cite{PhysRevD.107.103022}. As demonstrated in \cite{PhysRevD.109.083013}, this BGK subgrid model provides a good performance in reproducing macroscopic properties of FFC dynamics.

In the present study, we extend the BGK formalism to the three flavor framework. For that purpose we add a new idea. The point is that we need to treat competing three flavor-sectors simultaneously and the asymptotic state is unknown a priori as in the two flavor case. The key idea is that each sector is characterized with its own (instantaneous) relaxation time and asymptotic state individually. The asymptotic state of FFC in the three-flavor system $f^{a,3f}$ is then determined by solving the following \textit{ordinary} differential equations with respect to time for each angle,
\begin{equation}
    \begin{split}
        \frac{df_{e}}{dt}&=-\frac{f_{e}-f_{e}^{a,e\mu}}{\tau^{e\mu}}-\frac{f_{e}-f_{e}^{a,e\tau}}{\tau^{e\tau}},\\
        \frac{df_{\mu}}{dt}&=-\frac{f_{\mu}-f_{\mu}^{a,e\mu}}{\tau^{e\mu}}-\frac{f_{\mu}-f_{\mu}^{a,\mu\tau}}{\tau^{\mu\tau}},\\
        \frac{df_{\tau}}{dt}&=-\frac{f_{\tau}-f_{\tau}^{a,e\tau}}{\tau^{e\tau}}-\frac{f_{\tau}-f_{\tau}^{a,\mu\tau}}{\tau^{\mu\tau}}.
    \end{split}
    \label{eq:BGK_3f}
\end{equation}
One major advantage of this formulation is its ability to capture the competition between different flavor sectors in shaping the three-flavor asymptotic state.

A few important remarks regarding Eq.~\ref{eq:BGK_3f} are in order. First, $f^{a,ij}_{i/j}$ is determined from $f_i$ and $f_j$ at each time step, implying that $f^{a,ij}_{i/j}$ is also a time dependent quantity. This means that $f^{a,ij}_{i/j}$ does not represent an asymptotic state of flavor conversion, but rather it is a reference neutrino distribution, toward which the system evolves via FFC at each time. Secondly, we may use existing approximate methods for two neutrino flavor systems to determine both $f^{a,ij}_{i/j}$ and $\tau^{ij}$, whose details shall be described in Sec.~\ref{subsec:fandtau}. Thirdly, there are two relaxation terms on the right-hand side of Eq.~\ref{eq:BGK_3f}, which express the competition among three flavor sectors. Last but not least, despite its appearance, the BGK equation differs from the ordinary relaxation equation because its relaxation terms are inherently nonlinear in $f_i$, exhibiting an integral-equation nature through $\tau^{ij}$ and $f^{a,ij}_{i/j}$.

We now clarify the relationship between the three-flavor asymptotic state $f_{i}^{a,3f}$ and the sector-wise asymptotic state $f^{a,ij}_{i/j}$. Over the timescale of flavor conversions, $f_{i}^{a,3f}$ remains constant, whereas $f^{a,ij}_{i/j}$ can undergo significant evolution, as noted earlier. However, in the long term, the three-flavor asymptotic state $f_{i}^{a,3f}$, the two-flavor asymptotic states $f^{a,ij}_{i}$ and $f^{a,ik}_{i}$ (for $j\neq k$), and the neutrino distribution function $f_i$ will all evolve toward convergence, ultimately establishing a steady state. Thus we obtain the three-flavor asymptotic state by evolving Eq.~\ref{eq:BGK_3f} till a steady state.

One may wonder if we can directly determine $f_{i}$ satisfying that the right-hand side of Eq.~\ref{eq:BGK_3f} becomes zero. However, this approach is not appropriate to determine asymptotic states of FFCs, because the solution is not unique. This can be understood through a property of $\tau^{ij}$. As is well known, the relaxation time is determined based on a growth rate of flavor instability, indicating that $\tau^{ij}$ for FFC becomes infinity in cases with no angular crossings (see also \cite{Nagakura_2019,PhysRevResearch.2.012046}). On the other hand, there are, in general, an infinite number of solutions to eliminate the angular crossings for arbitrary initial neutrino distributions in all sectors, indicating that the solution can not be analytically derived. As we shall show in Sec.~\ref{subsec:results}, all $f_{i}^{a,3f}$ are determined by the interplay among all relaxation terms in the right-hand side of Eq.~\ref{eq:BGK_3f}, making analytic treatment impossible. Also, any other crude one-step estimation of $f_{i}^{a,3f}$ is inherently instantaneous, implying infinitely fast relaxation rates. Consequently, such an approach is incapable of accurately capturing the competition among different coherency sectors, each operating at distinct timescales. Consequently, directly solving Eq.~\ref{eq:BGK_3f} appears to be the only feasible approach to obtain these solutions. It should be mentioned that the numerical integration of Eq.~\ref{eq:BGK_3f} is very simple and computationally inexpensive, exhibiting no practical difficulties.

\subsection{Estimating $f^{a,ij}_{i/j}$ and $\tau^{ij}$}\label{subsec:fandtau} As already pointed out above, $f^{a,ij}_{i/j}$ and $\tau^{ij}$ can be computed based on existing approximate methods. In this subsection, we describe a recipe for them. For more details of our approximate methods, we refer readers to \cite{PhysRevD.107.103022,PhysRevD.109.083013}.

In the present study, we focus on FFC. In the fast limit, we ignore vacuum effects and matter refraction. We also neglect the collision term that leads to an energy dependence in FFC. We, hence, work in energy-integrated neutrino distributions throughout this paper. To estimate $f^{a,ij}_{i/j}$ and $\tau^{ij}$, we first define a relative angular distribution of neutrino lepton numbers between $i$ and $j$ flavors as
\begin{equation}  
G_{\mathbf{v}}^{\hspace{0.2mm}ij} \equiv \sqrt{2} G_{F} \bigg( \left(f_{i} - \bar{f}_{i}\right) - \left(f_{j} - \bar{f}_{j}\right) \biggl).
\label{eq:defG}
\end{equation}
Note that a factor of $\sqrt{2} G_F$, where $G_F$ denotes Fermi constant, is added in the definition of $G_{\mathbf{v}}^{\hspace{0.2mm}ij}$ by following a convention of \cite{PhysRevD.107.103022}. We then define its angular integrated quantities by dividing the integration domain into two regions by $G_{\mathbf{v}}^{\hspace{0.2mm}ij}$ as
\begin{equation}
    \begin{split}
     A &\equiv \left|\int_{G^{ij}_\mathbf{v} < 0} \! \frac{d\Omega}{4 \pi} \, G^{ij}_\mathbf{v} \right|, \\
     B &\equiv \left|\int_{G^{ij}_\mathbf{v} > 0} \! \frac{d\Omega}{4 \pi} \, G^{ij}_\mathbf{v} \right|.
    \end{split}
     \label{eq:def_AB}
\end{equation}
By using $A$ and $B$, we compute $\tau^{ij}$ as,
\begin{equation}
\tau^{ij} = \frac{2\pi} {\sqrt{AB}}.
\label{eq:taucomp}
\end{equation}
It should be mentioned that the absence of angular crossings implies either $A=0$ or $B=0$, which yields $\tau \to \infty$.

In our approximate scheme originally proposed by \cite{PhysRevD.107.103022}, $f^{a,ij}_{i/j}$ and $f^{a,ij}_{j}$ are computed through a survival probability of $i$ flavor of neutrinos ($P_{i,j}$) as,
\begin{equation}  
     \begin{split}  
       f^{a,ij}_{i/j} &= P_{i,j} f_{i} + (1 - P_{i,j}) f_{j}, \\  
       f^{a,ij}_{j} &= (1 - P_{i,j}) f_{i} + P_{i,j} f_{j}.  
     \end{split}  
\label{eq:faij_viaPi}
\end{equation} 
In computing $P_{i,j}$, $A$ and $B$ in Eq.~\ref{eq:def_AB} are also used. We assume a box-like angular distribution for $P_{i,j}$, and it can be computed as,
   \begin{equation}  
     P_{i,j} = \begin{cases}  
       \frac{1}{2} & \text{for } G^{ij}_\mathbf{v} < 0, \\  
       1 - \frac{A}{2B} & \text{for } G^{ij}_\mathbf{v} > 0.  
     \end{cases} 
     \label{eq:Pee1}
   \end{equation}
for cases with $B \ge A$, and
   \begin{equation}  
     P_{i,j} = \begin{cases}  
       \frac{1}{2} & \text{for } G^{ij}_\mathbf{v} > 0, \\  
       1 - \frac{B}{2A} & \text{for } G^{ij}_\mathbf{v} < 0.  
     \end{cases}  
     \label{eq:Pee2}
   \end{equation}
for the others. We note that Eq.~\ref{eq:faij_viaPi} ensures the total lepton number conservation with respect to each neutrino flight direction, while Eqs.~\ref{eq:Pee1}~and~\ref{eq:Pee2} also ensure the conservation of the zeroth angular moment (or number density) of $f_i - \bar{f}_i$. These are necessary conditions to develop a reliable approximate method for predicting asymptotic states of FFCs (see \cite{PhysRevD.107.103022} for more details).

\section{Validation}\label{sec:validation}
In this section, we test how well the new BGK asymptotic method can reproduce the results of quantum kinetic simulations. Hereafter, we work in the unit of $\mu\equiv\sqrt{2}G_\textbf{F}n_{\nu_e}$, so we omit $\mu$ in the following discussion unless otherwise stated.

\subsection{Models}\label{subsec:models}

\begin{table*}  
  \centering  
  \caption{Parameters for initial neutrino angular distribution in Eq.~\ref{eq:dist}. $ \xi_i $: Gaussian width; $ \alpha_i $: normalized number density. In each list, the parameters are given in the following order: $\nu_e, \bar{\nu}_e, \nu_\mu, \bar{\nu}_\mu, \nu_\tau, \bar{\nu}_\tau$.}  
  \label{tab:models}  
  \begin{tabular}{@{} l c c @{}}  
    \toprule  
    \textbf{Model} & $ (\xi_i) $ & $ (\alpha_i) $ \\  
    \midrule  
    Ef-2f & (0.6, 0.53, 0.5, 0.5, 0.5, 0.5) & (1.0, 0.92, 0.9, 0.9, 0.9, 0.9) \\  
    Cross1-3f & (0.6, 0.56, 0.52, 0.5, 0.3, 0.3) & (1.0, 0.92, 0.9, 0.82, 0.9, 0.9) \\  
    Cross2-3f & (0.6, 0.52, 0.45, 0.5, 0.3, 0.3) & (1.0, 0.92, 0.8, 0.9, 0.9, 0.9) \\  
    Cross3-3f & (0.6, 0.53, 0.48, 0.5, 0.3, 0.3) & (1.0, 0.92, 0.9, 0.9, 0.9, 0.9) \\  
    \bottomrule  
  \end{tabular}  
\end{table*}  

\begin{figure*}
    \centering
    \includegraphics[width=0.9\linewidth]{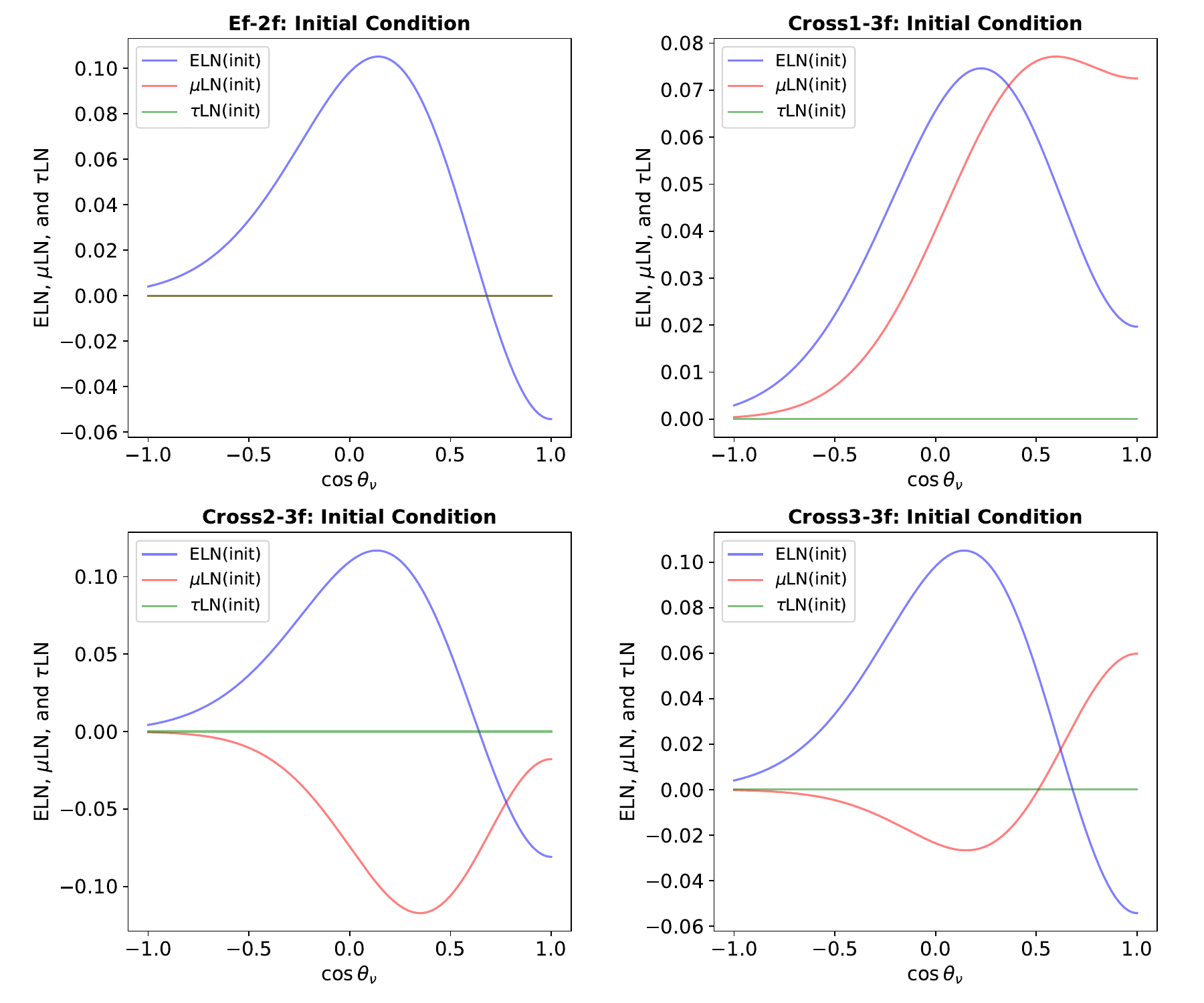}
    \caption{Initial conditions for models studied in this paper. Model parameters are given in Table.~\ref{tab:models} and distributions functions are given by Eq.~\ref{eq:dist}.}
    \label{fig:Init}
\end{figure*}

For all models, the initial conditions for neutrino angular distributions are set based on the following axially symmetric Gaussian function,
\begin{equation}  
  f_i = \frac{\alpha_i}{n_i}\exp\left[-\frac{(\cos\theta_\nu - 1)^2}{2\xi_{i}^2}\right],
\label{eq:dist}
\end{equation}  
where $n_i\equiv\int_{-1}^{1} \frac{dx}{2} \exp\left[-\frac{(x - 1)^2}{2\xi_{i}^2}\right]$ and $\theta_\nu$ is the angle with respect to the symmetry axis in the neutrino momentum space, which is the same as the spatial direction of the 1D box.
There are two control parameters in the function. $\xi_i$ characterizes the shape of the angular distribution and $\alpha_i$ represents the total amount (or angular-integrated) neutrino number density.

In this study, we set four models: Ef-2f, Cross1-3f, Cross2-3f, and Cross3-3f. The parameters for the initial condition of these models are summarized in Table~\ref{tab:models}. Their initial NFLN distributions are shown in Fig.~\ref{fig:Init}, where NFLN denotes neutrino flavor lepton number which is defined as $f_i-\bar{f}_i$.

Ef-2f model corresponds to an effective two-flavor system where $\mu$LN and $\tau$LN, denoting $\mu$ and $\tau$ neutrinos lepton number, respectively, are assumed to be identical. It should be noted that the asymptotic state in such a system can be predicted by an analytic scheme in \cite{PhysRevD.107.103022}. However, it is not a trivial issue whether our new method based on BGK prescription has the same level of performance as in \cite{PhysRevD.107.103022}. This is because their approaches are completely different from each other. As shown in Eqs.~\ref{eq:Pee1}~and~\ref{eq:Pee2}, the survival probability is set to be $1/2$ in cases with flavor equipartition, but it is $1/3$ for the analytic scheme in \cite{PhysRevD.107.103022}. The difference comes from the fact that the survival probability of $\nu_e$ is defined at each coherent sector for the BGK approach, but it is collectively treated in \cite{PhysRevD.107.103022}. It is necessary to show that our new BGK asymptotic method can achieve a precision similar to \cite{PhysRevD.107.103022} to validate its capability.

For the other models (Cross1-3f, Cross2-3f, and Cross3-3f), angular distributions of NFLN (which is hereafter denoted as lepton number) are different among all flavors (see Fig.~\ref{fig:Init}). For the Cross1-3f model, there is a crossing between ELN (which is defined as electron-type neutrinos lepton number) and $\mu$LN angular distributions. We note that $\tau$LN is assumed to be zero at all angles. For the Cross2-3f model, both ELN-$\mu$LN and ELN-$\tau$LN angular distributions show a crossing (but no crossings in $\mu$LN-$\tau$LN), indicating that all neutrino flavors should evolve with time. It is worthy of note that competing effects are involved in ELN distributions since both $e\mu$ and $e\tau$ sectors are unstable to FFIs. The Cross3-3f model corresponds to the most general case, in which a crossing is set in all flavor coherent sectors. This implies that FFC should occur in all of them. The intuitive example shown in Fig.~\ref{fig:3fpic} fundamentally aligns with the characteristics of this model. We will show in Sec.~\ref{subsec:results} that the BGK scheme can capture the qualitative trend even for such a complex system.

\begin{figure*}
    \centering
    \includegraphics[width=0.9\linewidth]{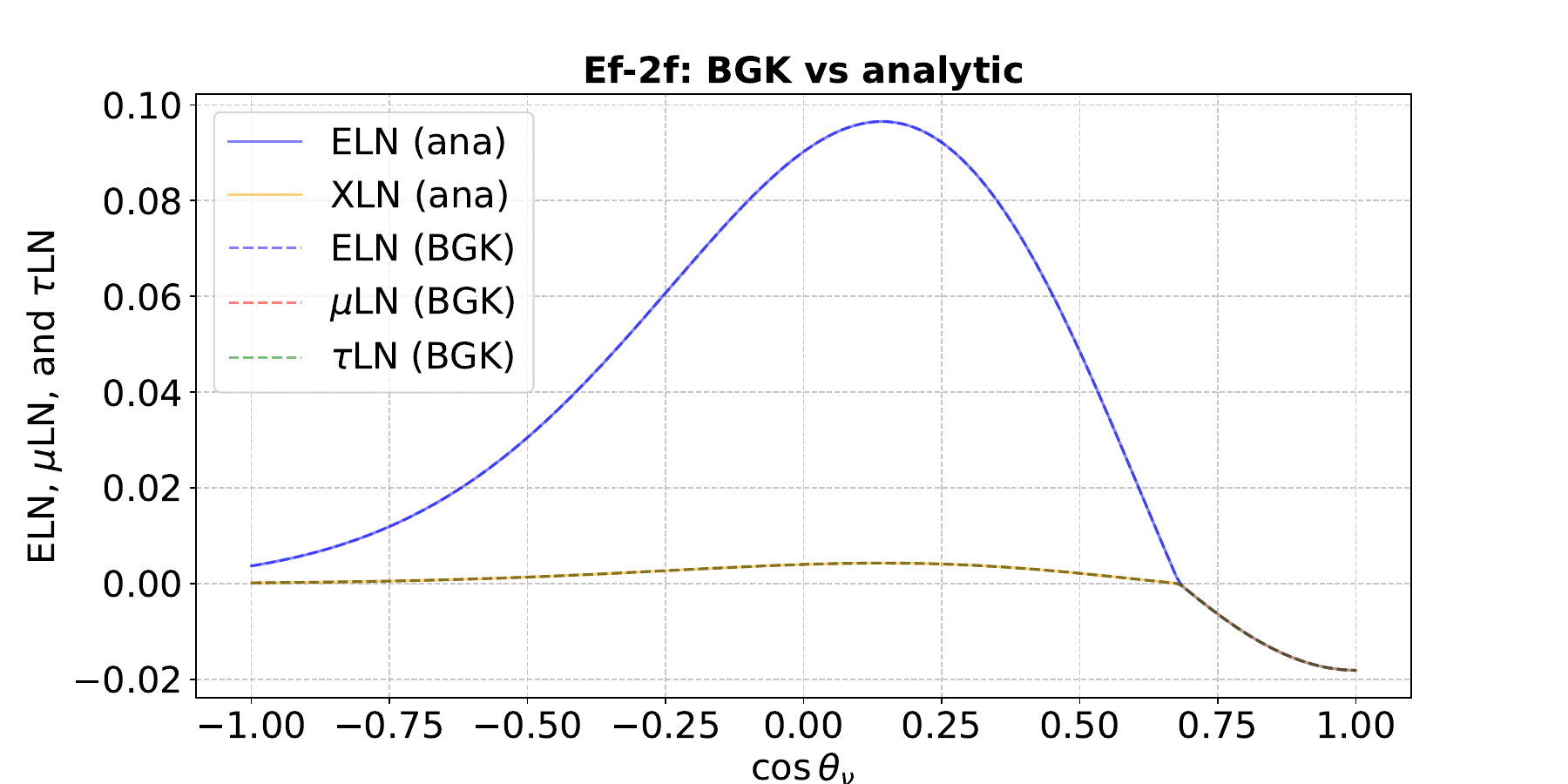}
    \caption{Asymptotic state for model Ef-2f given by our three-flavor BGK method (dashed lines) and an analytic scheme in \cite{PhysRevD.107.103022} (solid lines).
    }
    \label{fig:BGKvsANA}
\end{figure*}

\subsection{BGK and QKE solvers}\label{subsec:BGKQKEsolvers}

Before presenting our numerical results, let us briefly summarize our schemes for both BGK and QKE solvers. 
In our BGK asymptotic scheme, we solve Eq.~\ref{eq:BGK_3f} explicitly in time. The time step, $\delta t$, is set to be $10$ in the entire evolution, which is short enough to resolve flavor conversions for all models\footnote{The relaxation timescales approximated by Eq.~\ref{eq:taucomp} are at least of order $\mathcal{O}(10^2)\sim\mathcal{O}(10^3)$ in all models.}. It should be emphasized that $\tau^{ij}$ and $f_{i}^{a,ij}$ are computed based $f_i$ (see Sec.~\ref{subsec:fandtau}), which is updated at every time step. We also note that $\delta t$ should be varied at each time step for time-efficient computations when we implement it in subgrid models of flavor conversions (see discussions in Sec.~\ref{sec:conclusions}). These improvements are rather straightforward, which shall be done in our future work.

For quantum kinetic simulations, we employ a numerical code in \cite{PhysRevD.107.123021,PhysRevD.109.083031}. We solve QKE in one spatial dimension by seventh-order WENO method \cite{LIU1994200,GEORGE2023108588} with a fourth-order Runge-Kutta time integration (CFL number is set to be 0.4). The spatial domain is $L=1000$ and we impose a periodic boundary condition. We use a uniform grid of $N_z=10,000$ points in space. For angular direction in neutrino momentum space, it is covered with $ N_v = 256 $ Gaussian quadrature points. In all simulations, we assume axial symmetry in the neutrino momentum space.
At the beginning of the simulations, random perturbations in flavor coherency with magnitudes smaller than $10^{-6}$ times the diagonal elements are applied. In comparison to the BGK scheme, we take spatial averages of the neutrino distribution functions over the simulation box at the end of simulations when the system settles into a quasi-steady state.

\begin{figure*}
    \centering
    \includegraphics[width=0.9\linewidth]{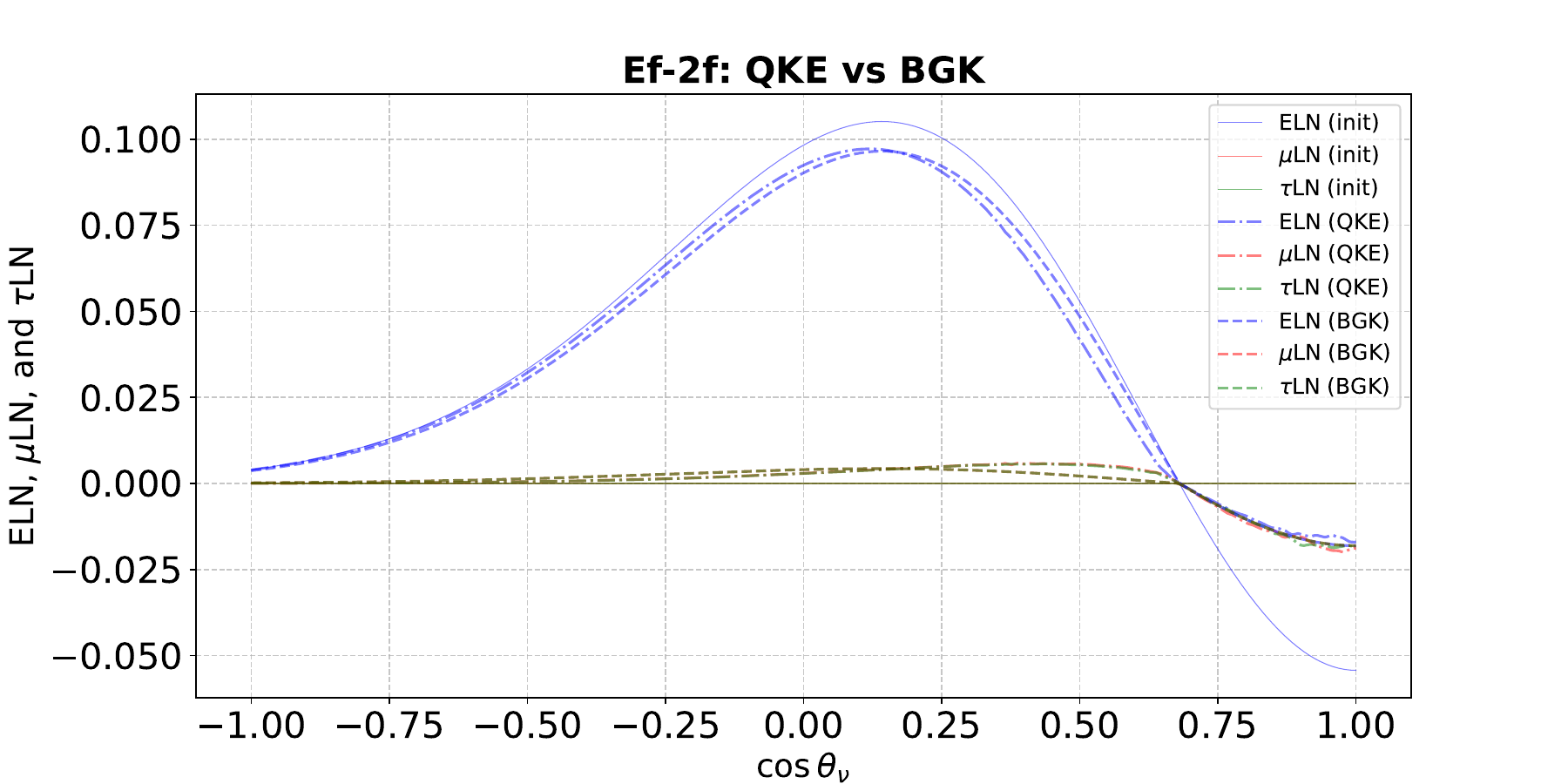}
    \caption{The initial ELN/$\mu$LN/$\tau$LN (blue/red/green) distribution functions (thin solid lines) and the asymptotic distribution functions derived by QKE simulations (dashed-dotted lines, extracted at $T_\text{QKE}=5000$) and by our three-flavor subgrid BGK scheme (dashed lines, extracted at $T_\text{BGK}=30000$). In model Ef-2f, the distribution functions for all heavy-leptonic (anti)neutrinos are identical.}
    \label{fig:asym_2f}
\end{figure*}

\subsection{Results}\label{subsec:results}

Before comparison with quantum kinetic simulations, we demonstrate that the BGK asymptotic method can reproduce the result from an analytic one in \cite{PhysRevD.107.103022} for Ef-2f model; see Fig.~\ref{fig:BGKvsANA}. As shown in the figure, the BGK method is in excellent agreement with the analytic one. It is worth noting that the angular region with $\cos{\theta_{\nu}} \gtrsim 0.7$ achieves flavor equipartition among all three flavors, even though we do not impose it in the BGK scheme. This result strengthens our basic assumption that spatially-averaged dynamics of FFC in the three-flavor system can be treated as a relaxation time process on multiple flavor coherency.

\begin{figure*}
    \centering
    \includegraphics[width=0.9\linewidth]{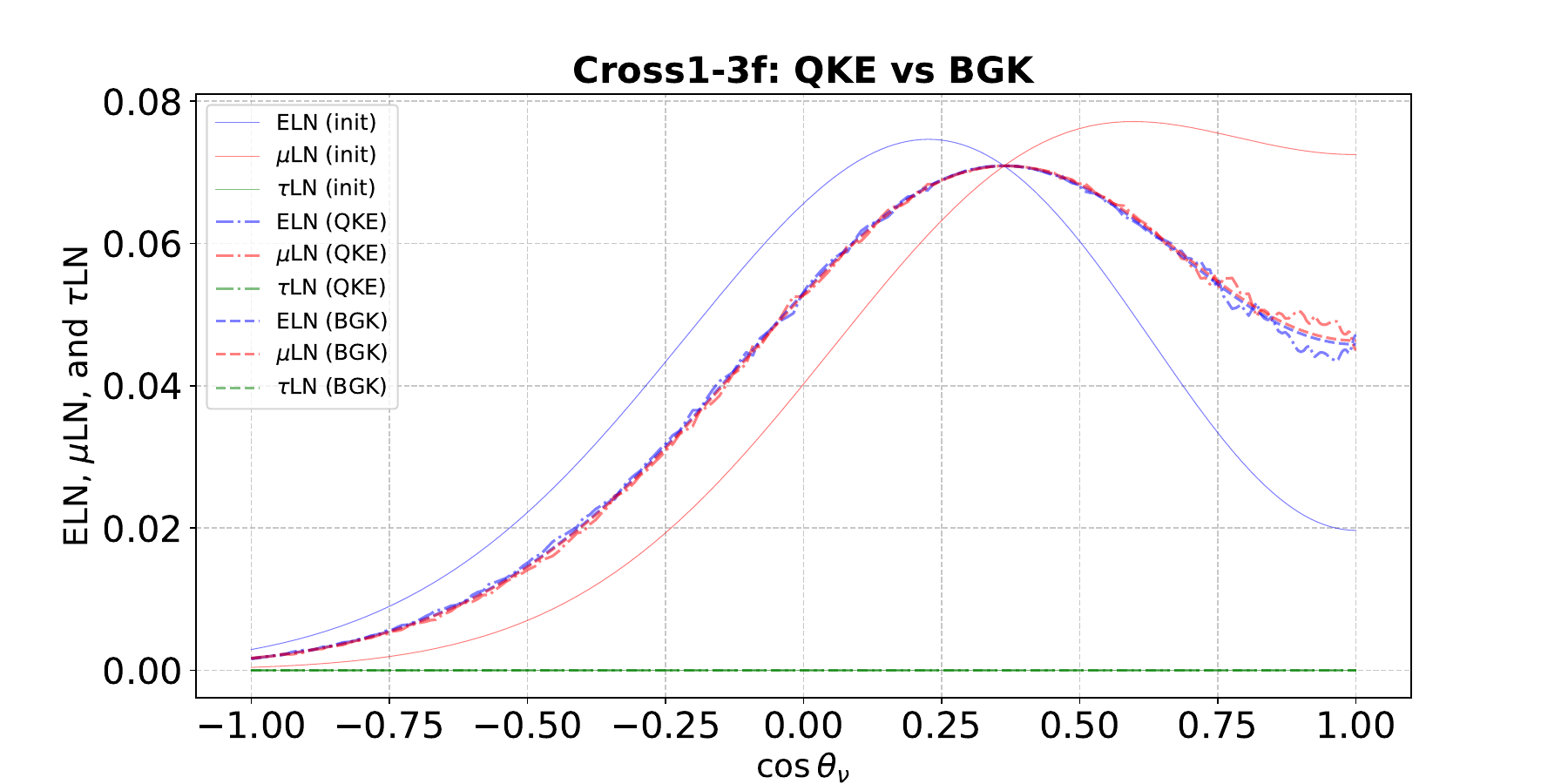}
    \caption{The same as Fig.~\ref{fig:asym_2f}, but for model Cross1-3f. Model Cross1-3f converges slower than others, and QKE simulation results are extracted at $T_\text{QKE}=15000$.}
    \label{fig:asym_Cross1-3f}
\end{figure*}

\begin{figure*}
    \centering
    \includegraphics[width=0.9\linewidth]{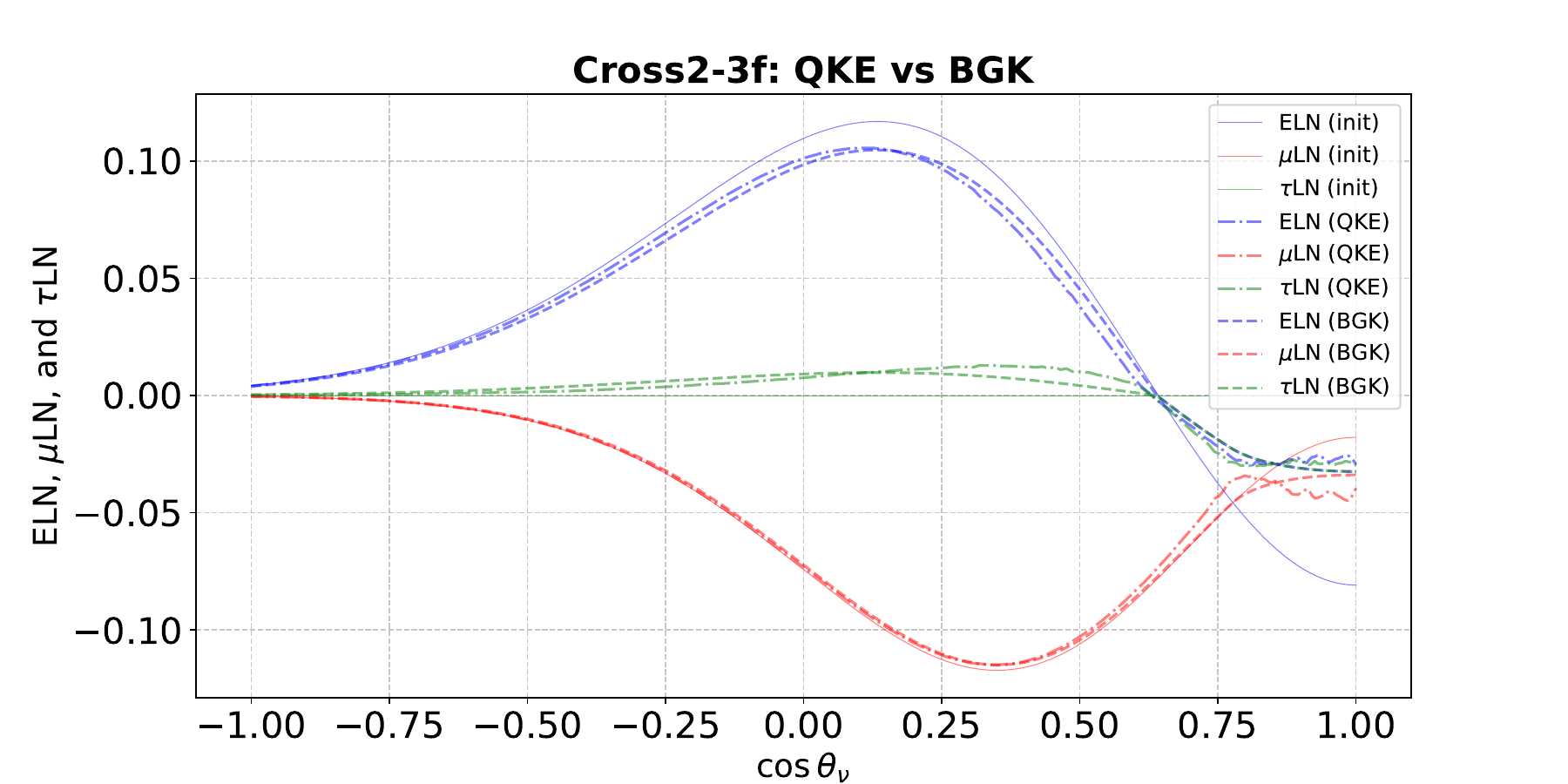}
    \caption{The same as Fig.~\ref{fig:asym_2f}, but for model Cross2-3f.}
    \label{fig:asym_Cross2-3f}
\end{figure*}

\begin{figure*}
    \centering
    \includegraphics[width=0.9\linewidth]{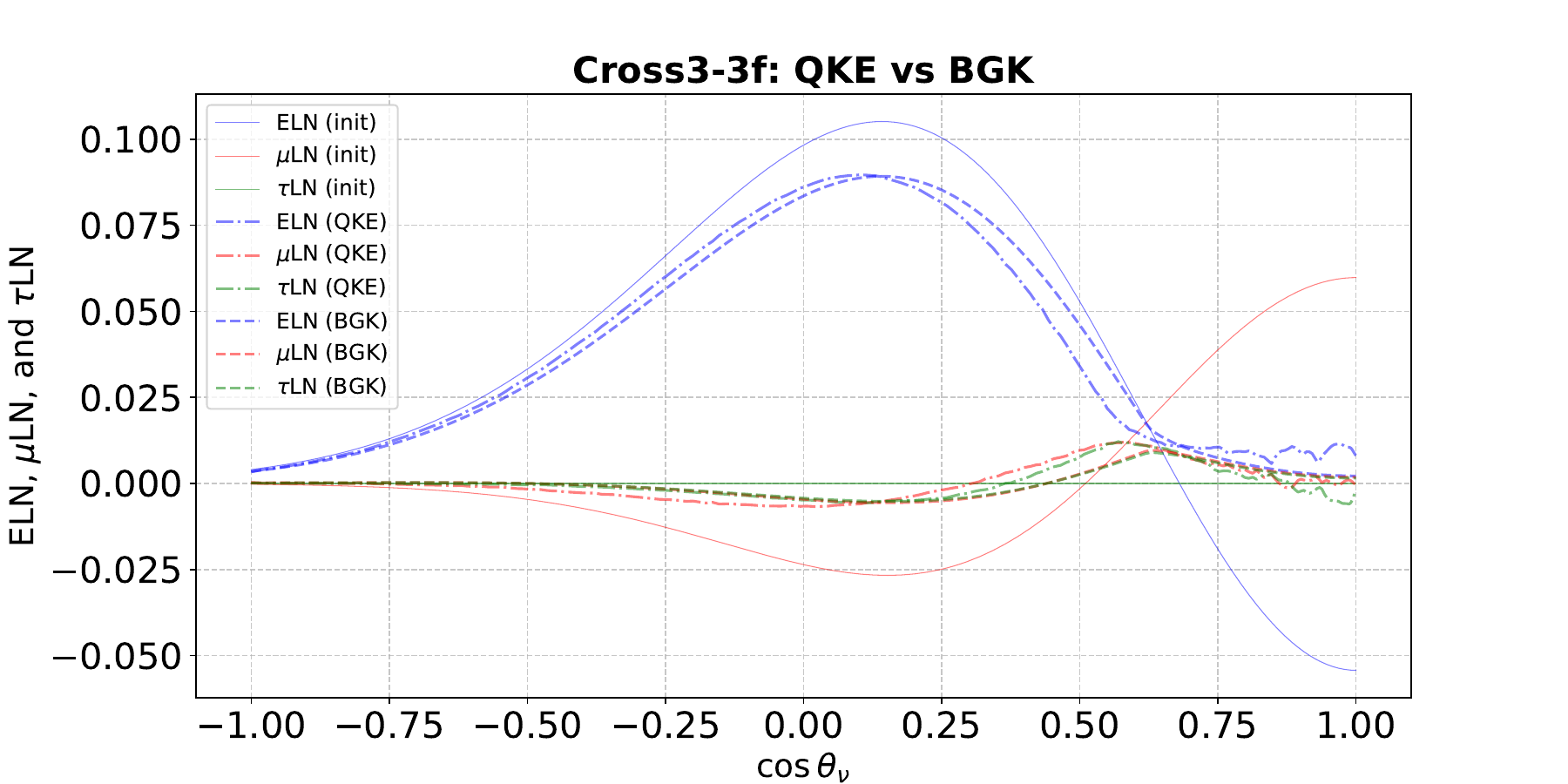}
    \caption{The same as Fig.~\ref{fig:asym_2f}, but for model Cross3-3f.}
    \label{fig:asym_Cross3-3f}
\end{figure*}

Let us turn our attention to comparison with quantum kinetic simulations. We show the asymptotic angular distributions of neutrinos in Figs.~\ref{fig:asym_2f}-\ref{fig:asym_Cross3-3f}. As a reference, the initial distributions of each model are also displayed with thin solid lines. Overall, we find that they are in good agreement with each other. One of the common features in these models is that angular crossings between two arbitrary NFLN distributions tend to reduce, which has also been observed in global simulations of FFCs in the literature (see, e.g., \cite{PhysRevLett.129.261101,PhysRevD.107.063033,PhysRevLett.130.211401,PhysRevD.108.123003,PhysRevD.109.123008}). In our BGK asymptotic scheme, the suppression of angular crossing is a general trend. This is because the system always evolves towards $f_{i}^{a,ij}$ with no angular crossings (see Sec.~\ref{subsec:fandtau}). One may wonder if FFCs of two sectors associated with one flavor can achieve a detailed balance with angular crossings, in which the right hand side of Eq.~\ref{eq:BGK_3f} vanishes. In this case, however, FFC in the other coherent sector must occur, which reduces the depth of angular crossings and then suppresses all crossings eventually. To facilitate the readers' understanding of this discussion, we provide a simple example and a mathematical argument in Appendix~\ref{appendix:balance}. In the following, we inspect each model more closely.

\begin{figure*}
    \centering
    \includegraphics[width=0.9\linewidth]{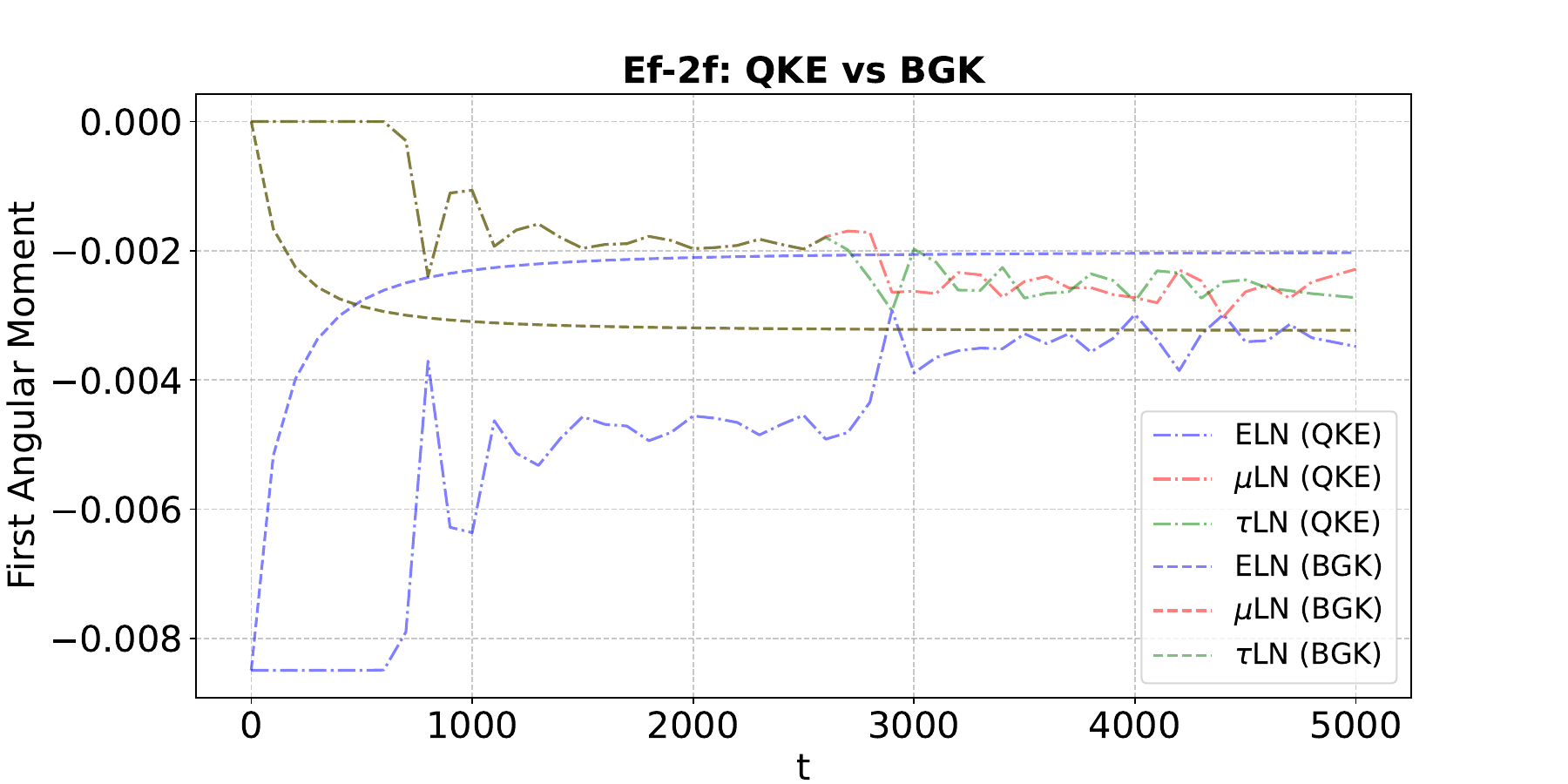}
    \caption{The time evolution of the first angular moment of ELN (blue lines), $\mu$LN (red lines), and $\tau$LN (green lines) in our BGK scheme (dashed lines) and numerical QKE simulations (dashed-dotted lines). }
    \label{fig:FAM_2f}
\end{figure*}

\begin{figure*}
    \centering
    \includegraphics[width=0.9\linewidth]{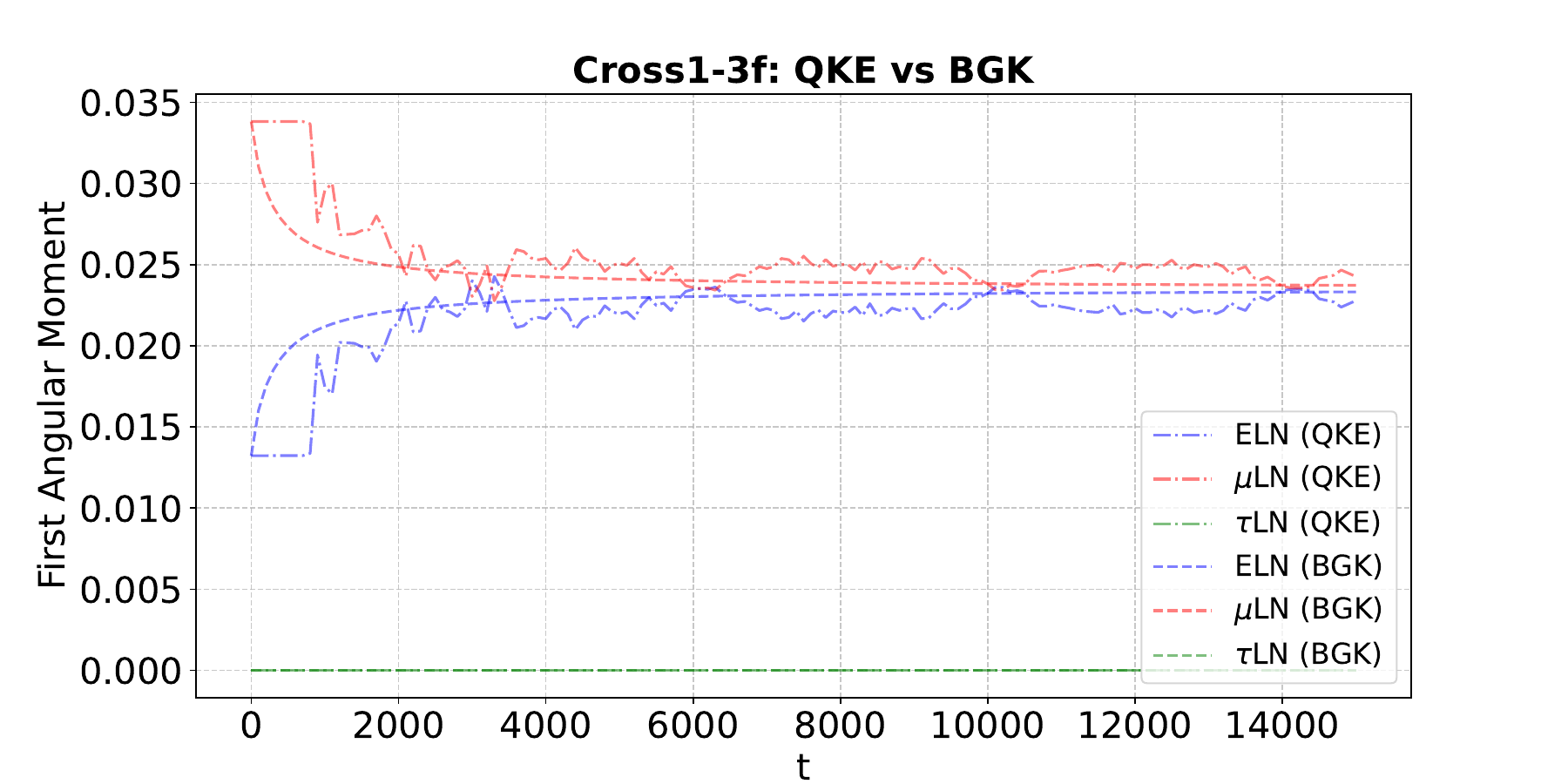}
    \caption{The same as Fig.~\ref{fig:FAM_2f}, but for model Cross1-3f.}
    \label{fig:FAM_Cross1-3f}
\end{figure*}

\begin{figure*}
    \centering
    \includegraphics[width=0.9\linewidth]{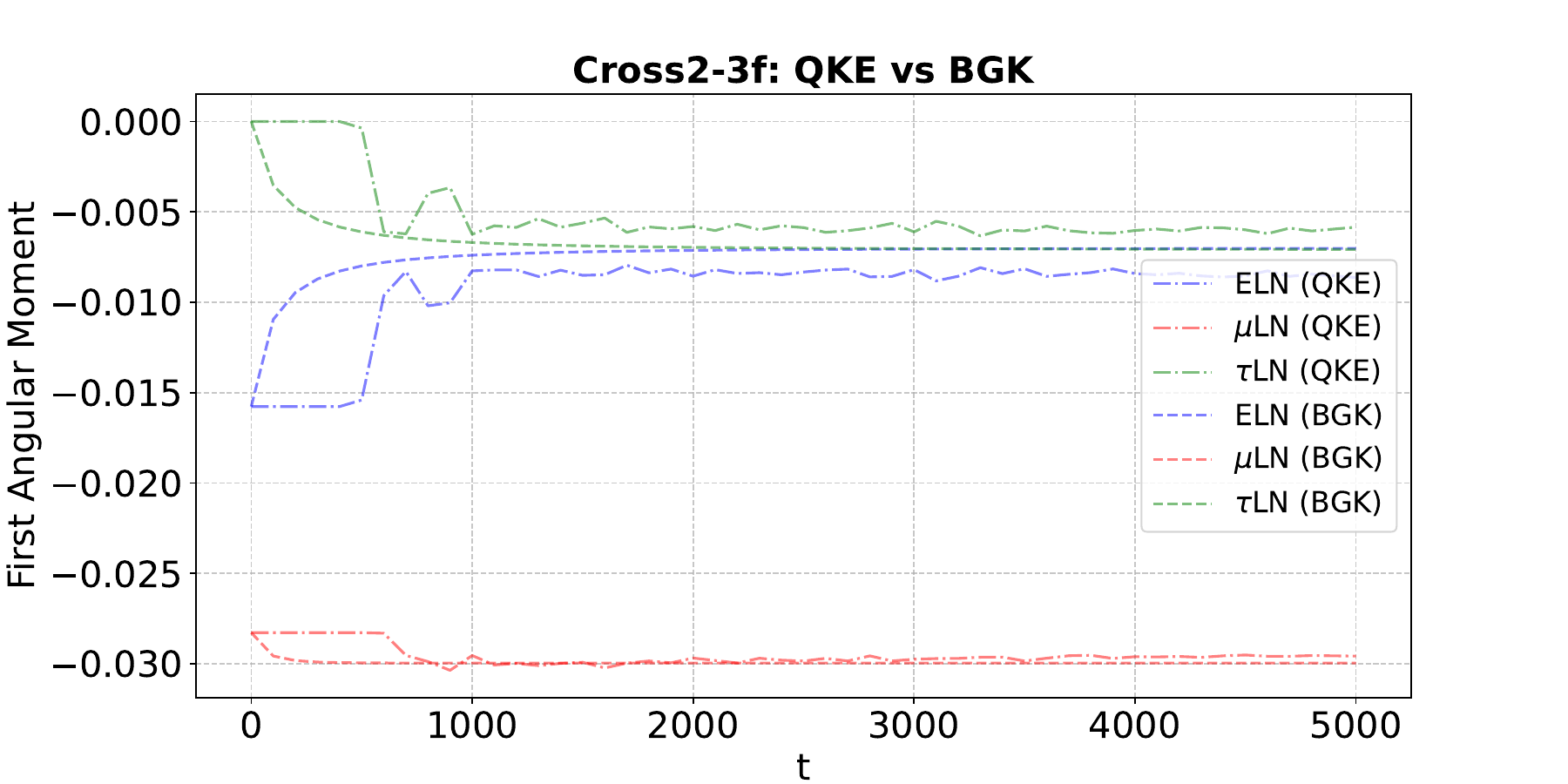}
    \caption{The same as Fig.~\ref{fig:FAM_2f}, but for model Cross2-3f.}
    \label{fig:FAM_Cross2-3f}
\end{figure*}

\begin{figure*}
    \centering
    \includegraphics[width=0.9\linewidth]{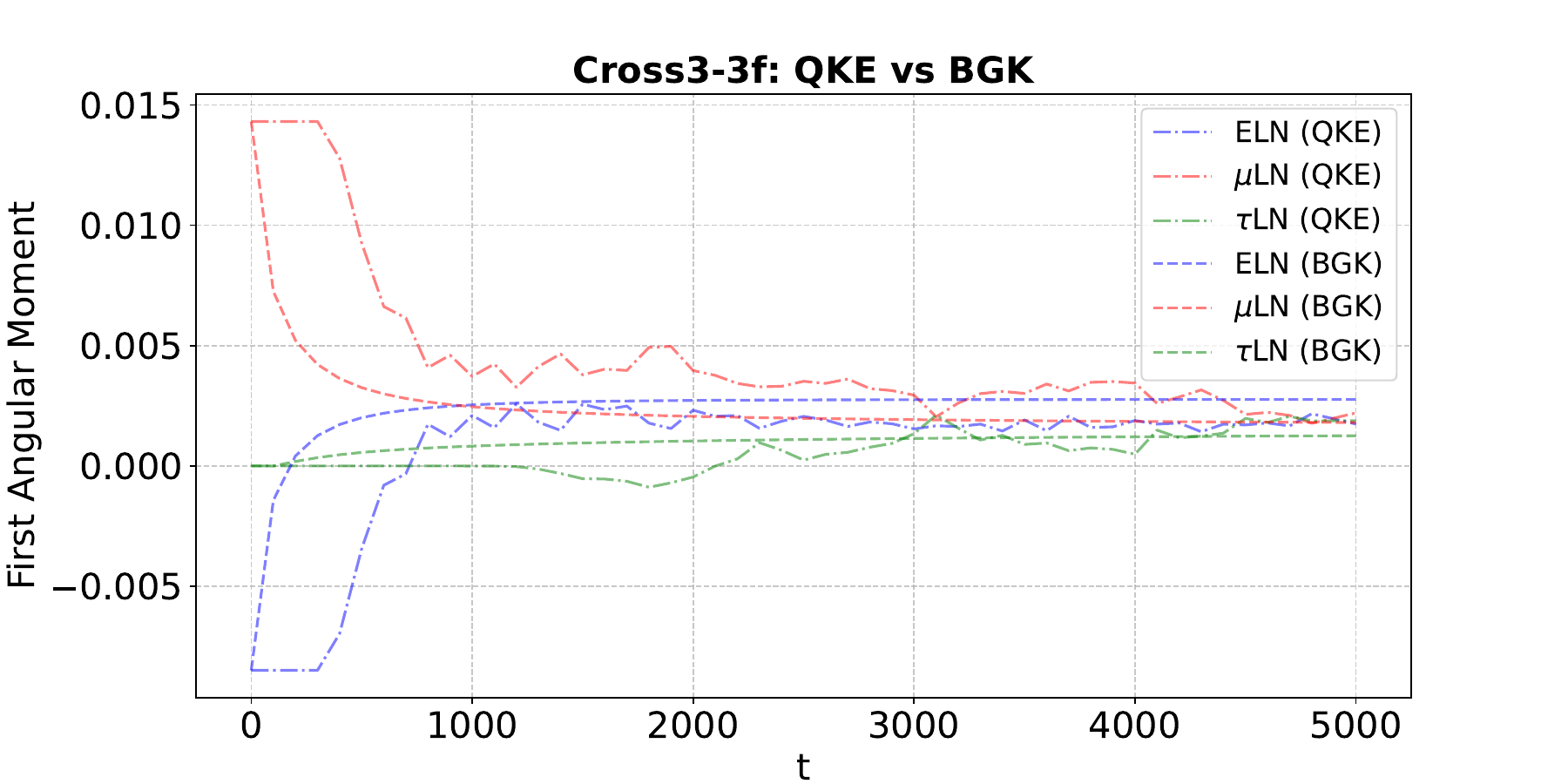}
    \caption{The same as Fig.~\ref{fig:FAM_2f}, but for model Cross3-3f.}
    \label{fig:FAM_Cross3-3f}
\end{figure*}

In Figs.~\ref{fig:FAM_2f}-\ref{fig:FAM_Cross3-3f}, we show the time evolution of the first angular moment of NFLN. For the result of quantum kinetic simulations, we take its spatial average over the entire simulation box. We note that the spatial average of the zeroth angular moment of NFLN is a conserved quantity in periodic boundary conditions (see \cite{PhysRevD.107.103022}). In fact, we use the conservation property to determine $f_{i}^{a,ij}$. For this reason, the zeroth angular moment is not adequate to assess the capability of the BGK approximate scheme. On the other hand, the first angular moment is a dynamical quantity and it corresponds to one of the representative quantities to show qualitative trends of angular distributions, suggesting that it is an appropriate physical quantity to check the capability of our BGK model.

For the Ef-2f model (see Fig.~\ref{fig:FAM_2f}), we confirm that both $\mu$LN and $\tau$LN have identical time evolutions in our BGK scheme. However, there is a deviation between the two flavors in quantum kinetic simulation, which is due to numerical noise. As discussed in \cite{PhysRevD.102.103017,PhysRevD.109.103040}, there is a turbulent-like property in FFC, implying that the small seed noise leads to large deviations in the nonlinear phase. On the other hand, the overall feature is statistically almost identical between the two flavors, so the numerical artifact does not compromise the present study.

Fig.~\ref{fig:FAM_2f} displays a remarkable deviation in the time evolution of the first angular moment between BGK and quantum kinetic simulations in the early phase. This error is mainly due to an approximate estimation of the relaxation time scale (see Eq.~\ref{eq:taucomp}). As discussed in \cite{PhysRevD.109.083013}, the empirical formula to estimate the growth rate of fast instabilities has only an ability to provide an order of magnitude estimation, exhibiting that it has limited quantitative accuracy. We also find that some complex features in the nonlinear FFC phase ($1000 \lesssim t \lesssim 3000$) are missed in the BGK model. The deviation is mainly due to large temporal variations appearing in the nonlinear phase of FFCs. They are driven by local ELN-$\mu$($\tau$)LN angular crossings that occur even after the system enters a quasi-steady state. Let us emphasize that these detailed features of FFCs are not our interest in subgrid models, and it is not surprising to see this level of the deviation. On the other hand, the error of asymptotic states of flavor conversion in our BGK method is within a few tens of percent, showing a good performance of our BGK method. We note that the major source of error is not due to the BGK approximation but rather to a boxlike treatment for angular distributions of survival probability of neutrinos. This indicates that errors can be reduced by adopting more elaborate methods as proposed in \cite{PhysRevD.108.063003}.

In the Cross1-3f model (see Fig.~\ref{fig:FAM_Cross1-3f}), we find that both the BGK method and quantum kinetic simulations show no time evolution of the first angular moment of $\tau$LN. It should be noted that the quantum kinetic simulation shows that the flavor coherency in $\nu_{e}-\nu_{\tau}$ and $\nu_{\mu}-\nu_{\tau}$ are six orders of magnitude smaller than that in $\nu_{e}-\nu_{\mu}$, exhibiting no growth of FFC in the two flavor-mixing sectors involving $\tau$ flavor. For the time evolution of ELN and $\mu$LN, the overall trend is well captured by our BGK method.

In the Cross2-3f model (see Fig.~\ref{fig:FAM_Cross2-3f}), it is worth noting that ELN and $\tau$LN evolve faster than $\mu$LN, and the amount of change of ELN and $\tau$LN in the early phase is also larger than $\mu$LN. This is because the angular crossing in ELN-$\tau$LN is deeper than in ELN-$\mu$LN at the initial condition, which causes the most vigorous FFCs for the ELN-$\tau$LN sector. As shown in Fig.~\ref{fig:asym_Cross2-3f}, the ELN and $\tau$LN almost achieve flavor equipartition between the two flavors in the angular region of $\cos {\theta_{\nu}} \gtrsim 0.75$. As a result, the depth of angular crossing in ELN-$\mu$LN becomes shallower, reducing flavor conversions between ELN and $\mu$LN. In fact, $\mu$LN does not reach a flavor equipartition in the angular region of $\cos{\theta_{\nu}} \gtrsim 0.75$. The BGK method successfully captures such a dynamic interaction of flavor conversions among all three different flavor coherent states.

\begin{figure*}
    \centering
    \includegraphics[width=0.9\linewidth]{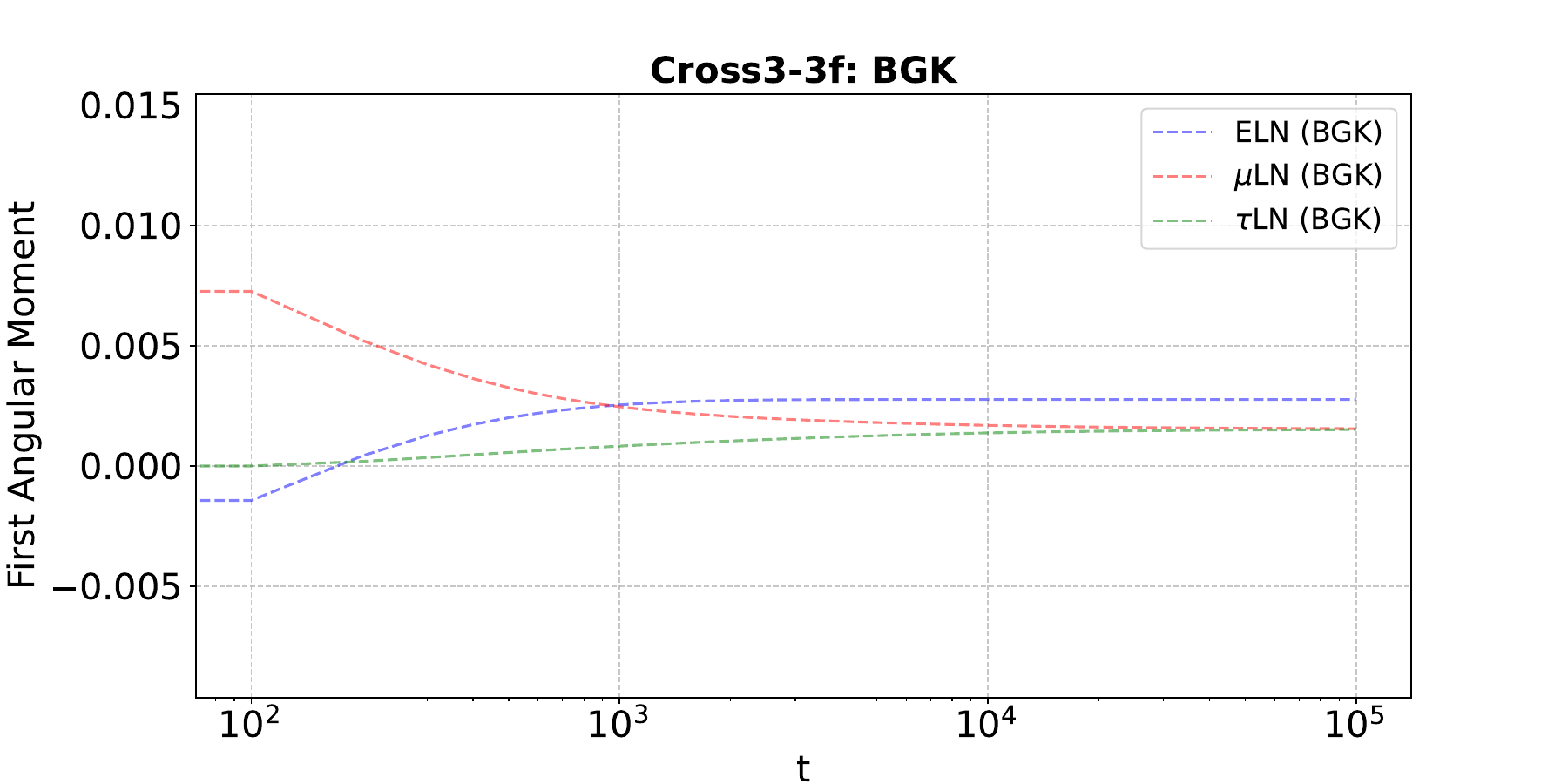}
    \caption{The result of the extended time evolution of BGK asymptotic simulation for the Cross3-3f model. We confirm that $\mu$LN and $\tau$LN appear to converge at around $t\gtrsim30000$.}
    \label{fig:FAM_Cross3-3f_LONG}
\end{figure*}

Finally, Fig.~\ref{fig:FAM_Cross3-3f} displays the result of the Cross3-3f model. As can be seen in the figure, fast instabilities occur in all flavor coherent sectors, which makes the overall dynamics more complicated than in other models; nevertheless, the BGK model shows a good performance. Another noticeable feature displayed in Fig.~\ref{fig:FAM_Cross3-3f} is that the first angular moment of $\mu$LN and $\tau$LN converge towards the same value for the BGK model. This convergence is more clearly illustrated in Fig.~\ref{fig:FAM_Cross3-3f_LONG}, where it is evident at $t\gtrsim30000$. This trend is consistent with their angular distributions in asymptotic states (see Fig.~\ref{fig:asym_Cross3-3f}), in which $\mu$LN and $\tau$LN have identical angular distributions. The result is expected from our choice of parameters for the initial condition, in which zeroth angular moment, $n$, for $\mu$LN and $\tau$LN, are set to be $n_{\rm \mu LN}=n_{\rm \tau LN} (=0)$ (see also $\alpha$ in Table~\ref{tab:models}), while they are conserved quantities during the time evolution. In this case, the survival probability, $P_{i,j}$ for the $\mu$LN-$\tau$LN sector becomes $1/2$ for all angles (see Eqs.~\ref{eq:Pee1}~and~\ref{eq:Pee2}), resulting in flavor equipartition between the two flavors. We note, on the other hand, $n_{\rm ELN}$ is positive, which leads to inhomogeneous angular distributions of survival probability. As a result, the asymptotic angular distribution of ELN is clearly different from other flavors.

\section{Implementation in subgrid model}\label{sec:impsubgrid}
In this section, we discuss how the BGK asymptotic scheme presented in this paper can be implemented in a classical neutrino transport code. One might think that the basic equation can be given by adding advection terms in Eq.~\ref{eq:BGK_3f}, i.e.,
\begin{equation}
    \begin{split}
         \left(\partial_t+\boldsymbol{v}\cdot\nabla\right) f_{e}&=-\frac{f_{e}-f_{e}^{a,e\mu}}{\tau^{e\mu}}-\frac{f_{e}-f_{e}^{a,e\tau}}{\tau^{e\tau}},\\
         \left(\partial_t+\boldsymbol{v}\cdot\nabla\right) f_{\mu}&=-\frac{f_{\mu}-f_{\mu}^{a,e\mu}}{\tau^{e\mu}}-\frac{f_{\mu}-f_{\mu}^{a,\mu\tau}}{\tau^{\mu\tau}},\\
         \left(\partial_t+\boldsymbol{v}\cdot\nabla\right) f_{\tau}&=-\frac{f_{\tau}-f_{\tau}^{a,e\tau}}{\tau^{e\tau}}-\frac{f_{\tau}-f_{\tau}^{a,\mu\tau}}{\tau^{\mu\tau}}.
    \end{split}
    \label{eq:BGKsubg_3f_v1}
\end{equation}
Although this can certainly serve as a subgrid-transport equation, we do not recommend using it; the reason is as follows. As discussed in the above sections, both $f_{i}^{a,ij}$ and $\tau^{ij}$ are time-dependent quantities, and their dynamical timescales are the order of those of fast instabilities. This suggests that the timestep of the simulation needs to be short enough to resolve FFCs, otherwise competing effects of FFC occurring in multiple flavor coherent sectors can not be handled precisely. However, there is a vast difference in scales between flavor conversions and astrophysical phenomena, indicating that it is not useful for global simulations.

Due to this issue, we propose another way. We adopt the original formulation of the BGK subgrid model; see Eq.~\ref{eq:BGK_flat}. In this formulation, we use $f_{i}^{a,3f}$, which is different from $f_{i}^{a,ij}$. More concretely, we solve Eq.~\ref{eq:BGK_3f} independently from Eq.~\ref{eq:BGK_flat} to determine $f_{i}^{a,3f}$ at each time snapshot and each spatial position (mesh). The relaxation time scale, $\tau$, can also be estimated as the elapsed time until the system reaches a quasi-steady state, or as several times larger, provided it remains shorter than the transport timescale. By using the $f_{i}^{a,3f}$ and $\tau$, we can solve Eq.~\ref{eq:BGK_flat} with the same order of timestep as that used in classical neutrino transport.

It should be noted, however, that we need to carry out a systematic study of the BGK subgrid model with global neutrino transport simulations in order to check the validity of the proposed method. Although these detailed investigations are beyond the scope of this paper, they are currently underway and will be presented in future work.

\section{Conclusions}\label{sec:conclusions}
In the standard model of CCSN and BNSM, $\mu$- and $\tau$ neutrinos and their antipartners have been assumed to be identical and therefore have been treated collectively. In studies of collective neutrino oscillations, we usually followed the convention. The recent theoretical indication that stable muons can appear in dense neutrino environments such as the vicinity of neutron stars exhibits a clear limitation in these previous studies. Motivated by this fact, we tackle an important issue of how to predict asymptotic states of FFCs in the three-flavor framework. Addressing this issue, however, requires taking into account competing effects of flavor conversions driven by multiple flavor coherency, which can not be handled analytically (see Sec.~\ref{sec:nontrivial}).

In Sec.~\ref{sec:BGKmethod}, we provide our basic strategy to determine asymptotic states of FFCs for arbitrary neutrino angular distributions. The key assumption in this approach is that we decouple flavor conversions in each flavor coherent sector. We then apply a BGK prescription for each sector, in which neutrinos are relaxed to a certain distribution (which is denoted as $f_{i}^{a,ij}$). For each flavor, there are two associated flavor coherent states, and their competing effects can be handled as shown in the right-hand side of Eq.~\ref{eq:BGK_3f}. It should be mentioned that both $f_{i}^{a,ij}$ and $\tau^{ij}$ are determined from $f_{i}$, implying that they are also time-dependent quantities. We, hence, solve Eq.~\ref{eq:BGK_3f} numerically, and then approximately estimate the asymptotic state of flavor conversions, $f_{i}^{a,3f}$ at $t \gg 1$. Although solving Eq.~\ref{eq:BGK_3f} requires numerical integration, it is computationally cheap and can be implemented easily in any classical neutrino transport codes (see Sec.~\ref{sec:impsubgrid}), and its numerical recipe of how to compute $f_{i}^{a,ij}$ and $\tau^{ij}$ is provided in Sec.~\ref{subsec:fandtau}. This procedure is essentially the same as existing analytic methods (see, e.g., \cite{PhysRevD.107.103022,PhysRevD.109.083013}), exhibiting that the implementation of this new BGK asymptotic scheme into classical neutrino transport should be straightforward.

In Sec.~\ref{subsec:results}, we demonstrate the ability of the BGK asymptotic scheme by comparing it to quantum kinetic simulations. In this test, we systematically change the initial neutrino angular distributions by controlling two independent parameters characterizing initial neutrino angular distributions (see Eq.~\ref{eq:dist} and Table~\ref{tab:models}). From the Ef-2f model, in which $\mu$LN and $\tau$LN are assumed to be identical, we confirm that the BGK asymptotic scheme reproduces the result of the analytic scheme by \cite{PhysRevD.107.103022} as well as quantum kinetic simulations. For Cross1-3f, Cross2-3f, and Cross3-3f models, we find good agreement between the BGK method and quantum kinetic simulations. In the same section, we also provide discussions about the source of errors in the BGK scheme.

In the present study, we focus on asymptotic states of FFC, but the BGK formalism, Eq.~\ref{eq:BGK_3f}, can be used for any flavor conversions associated with other types of instabilities. One of the interesting applications would be for collisinal flavor instability (CFI) \cite{PhysRevLett.130.191001}. The instability would also ubiquitously occur in CCSN \cite{PhysRevD.107.083016,PhysRevD.108.123024,PhysRevD.109.023012} and BNSM environments \cite{PhysRevD.108.083002}. As shown in a previous study \cite{PhysRevD.110.043039}, the appearance of stable muons affects occurrences of CFI. This also suggests that flavor conversions would undergo a complex interplay among different flavor coherency in the nonlinear phase. Our BGK asymptotic scheme can handle it, although we need to consider how $\tau^{ij}$ and $f_{i}^{a,ij}$ can be predicted for CFI. For the relaxation timescale, we can use an analytic estimation of the growth rate of CFI \cite{PhysRevD.107.123011}. However, no reasonable approximations to determine the asymptotic state have been proposed in the literature. Very recently, on the other hand, important progress has been made in \cite{zaizen2025spectraldiversitycollisionalneutrinoflavor}, in which the asymptotic states of CFI could be characterized by plus- and minus instability modes. We will address this important issue in future work.
 
As presented in this paper, subgrid models of flavor conversions can be improved step-by-step toward practical implementation into classical neutrino transport. This could be a key to understanding CCSN and BNSM. In fact, evidence has accumulated that flavor conversions occur in these environments \cite{PhysRevD.104.083025,Nagakura_2019,PhysRevD.109.023012,PhysRevD.103.063033,PhysRevD.106.083005,PhysRevD.108.083002,PhysRevD.107.083016,Mukhopadhyay_2024} and that they play substantial roles on both CCSN \cite{PhysRevLett.130.211401,PhysRevD.108.123003,PhysRevD.107.103034,PhysRevLett.131.061401,PhysRevD.109.123008,mori2025threedimensionalcorecollapsesupernovamodels,shalgar2025neutrinoquantumkineticsflavors} and BNSM dynamics \cite{PhysRevLett.126.251101,PhysRevD.106.103003,PhysRevD.105.083024}.

\begin{acknowledgments}
H.N. is supported by Grant-in-Aid for Scientific Research (23K03468). 
L.J. is supported by a Feynman Fellowship through LANL LDRD project number 20230788PRD1. 
M.Z. is supported by Grant-in-Aid for JSPS Fellows (Grant No. 22KJ2906) and JSPS KAKENHI Grant Number JP24H02245. 
S.Y. is supported by the Institute for Advanced Theoretical and Experimental Physics, Waseda University, and the Waseda University Grant for Special Research Projects (project No. 2024C-56, 2024Q-014).
Numerical computation in this work was partly carried out at the Yukawa Institute Computer Facility. 
\end{acknowledgments}

\bibliography{refe}
\bibliographystyle{apsrev4-2}

\appendix
\section{No asymptotic states of FFCs with ELN-$\mu$($\tau$)LN angular crossings} \label{appendix:balance}
\begin{figure*}
    \centering
    \includegraphics[width=1\linewidth]{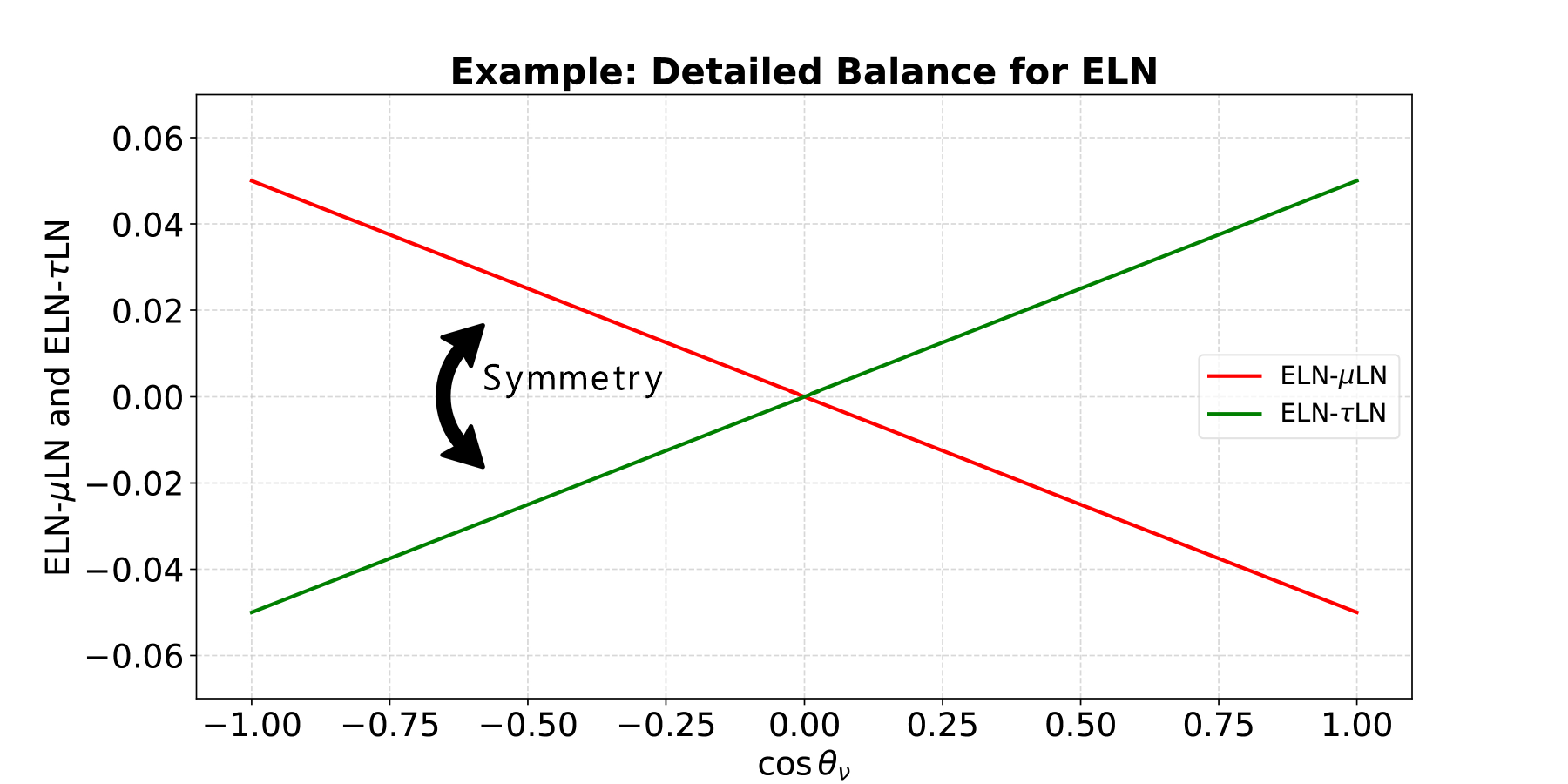}
    \caption{Symmetric ELN-$\mu$LN and ELN-$\tau$LN distributions with respect to $y=0$, where $y$ denotes ELN-$\mu$LN or ELN-$\tau$LN. In this case, ELN is in a detailed balance state (see Eq.~\ref{eq:BGK_3f}). However, there exists a zero crossing in $\mu$LN-$\tau$LN=(ELN-$\tau$LN)-(ELN-$\mu$LN), which leads to fast instabilities in the $\mu\tau$ sector. This instability will merge $\mu$LN and $\tau$LN, whose asymptotic states are exactly the ELN distribution. See text for more details.}
    \label{fig:balance}
\end{figure*}  
We provide a concrete example for our claim in Sec.~\ref{subsec:results}, that there are no such asymptotic states of FFCs achieving detailed balance among all flavors with angular crossings still existing. Suppose that ELN-$\mu$LN and ELN-$\tau$LN angular distributions are exactly symmetric (see Fig.~\ref{fig:balance}). In this case, our BGK scheme implies $df_{e}/dt=0$ due to achieving a detailed balance between the two relaxation terms associated with ELN in Eq.~\ref{eq:BGK_3f}. On the other hand, the following relation, (ELN-$\mu$LN)-(ELN-$\tau$LN)=$\tau$LN-$\mu$LN, implies that crossings should exist between $\tau$LN and $\mu$LN angular distributions. As a result, FFC occurs in the $\mu-\tau$ sector until the crossings disappear. Due to the symmetry condition, the effective two-flavor FFC in the $\mu-\tau$ sector ensures that $(\rm{ELN-\mu LN})-(\rm{ELN-\tau LN})=\rm{\tau LN}-\rm{\mu LN}=0$ in the asymptotic state. As a result, we obtain $\rm{\tau LN}=\rm{\mu LN}$, implying that $\mu$LN and $\tau$LN achieve flavor equipartition, which exactly coincide with the ELN distribution in all angles as their asymptotic states. This example exhibits that the symmetric initial condition between two NFLN angular distributions with respect to the other NFLN always leads to flavor equipartition for all three NFLN as their asymptotic states.

We can also provide more mathematical proof regarding this argument. Assume that three NFLN angular distributions achieve detailed balances with crossings (i.e., symmetric distributions similar to Fig.~\ref{fig:balance}). In such situations, there are angular regions that satisfy $\tau$LN$>$ELN$>$$\mu$LN (or $\mu$LN$>$ELN$>$$\tau$LN). On the other hand, the condition of detailed balance between $\mu$LN-ELN and $\mu$LN-$\tau$LN implies that ELN$>$$\mu$LN$>$$\tau$LN (or $\tau$LN$>$$\mu$LN$>$ELN) should also be satisfied. However, there are no solutions that satisfy both conditions simultaneously, implying that there are no NFLN angular distributions with detailed balance among the three flavors.
\end{document}